\documentclass[twocolumn,pra,superscriptaddress,showpacs,amsmath]{revtex4}

\usepackage{graphicx}
\usepackage{epsfig}
\usepackage{graphics}
\begin{document}
\title{Strong magnetic coupling of an ultracold gas to a \\ superconducting waveguide cavity}

\author{J. Verd\'{u}}
\affiliation{Atominstitut der \"Osterreichischen Universit\"aten, TU-Wien, 1020 Vienna, Austria}
\author{H. Zoubi}
\affiliation{Institut f\"{u}r Theoretische Physik, Universit\"{a}t Innsbruck, Technikerstr. 21a, 6020 Innsbruck, Austria}
\author{Ch. Koller}
\affiliation{Atominstitut der \"Osterreichischen Universit\"aten, TU-Wien, 1020 Vienna, Austria}
\author{J. Majer}
\affiliation{Atominstitut der \"Osterreichischen Universit\"aten, TU-Wien, 1020 Vienna, Austria}
\author{H. Ritsch}
\affiliation{Institut f\"{u}r Theoretische Physik, Universit\"{a}t Innsbruck, Technikerstr. 21a, 6020 Innsbruck, Austria}
\author{J. Schmiedmayer}
\affiliation{Atominstitut der \"Osterreichischen Universit\"aten, TU-Wien, 1020 Vienna, Austria}

\date{\today}

\begin{abstract}
Placing an ensemble of $10^6$ ultracold atoms in the near field of
a superconducting coplanar waveguide resonator (CPWR) with $Q \sim
10^6$ one can achieve strong coupling between a single microwave
photon in the CPWR and a collective hyperfine qubit state in the
ensemble with $g_\textit{eff} / {2 \pi} \sim 40$ kHz larger than
the cavity line width of ${\kappa }/{2 \pi} \sim 7$ kHz.
Integrated on an atomchip such a system constitutes a hybrid
quantum device, which also can be used to interconnect solid-state
and atomic qubits, to study and control atomic motion via the
microwave field, observe microwave super-radiance, build an
integrated micro maser or even cool the resonator field via the
atoms.
\end{abstract}


\maketitle

In the past decade important breakthroughs in implementing quantum
information processing were reached in different physical
implementations \cite{Bouwmeester:2000}, each showing advantages
and shortcomings. For quantum information to emerge as a valuable
technology, it is mandatory to pool their strengths. Solid-state
systems allow fast processing and dense integration; atom or ion
based systems are slower but exhibit long qubit coherence times.
Ensembles of atoms constitute a  quantum memory, it can be read
out onto photons \cite{Lukin:2003} which can then be transmitted
over long distances \cite{Gisin:2002}. Here we analyze a device to
\emph{quantum interconnect} superconducting solid-state qubits to
an atomic quantum memory.

The challenge in transferring the state of a solid-state qubit to
atoms is bridging the tremendous gap in time scales that govern
solid-state and atomic physics devices. This difference can be
overcome using a coplanar waveguide resonator (CPWR)
\cite{Day:2003,Mazin:2004,Frunzio:2005g}, which can be
electrically coupled to single superconducting qubits
\cite{Blais:2004,Wallraff:2004,Sillanpaa:2007,Hofheinz:2008,Fink:2008}.
Various ways were proposed to couple to atomic and molecular
systems
\cite{Sorensen:2004,Tian:2004,Andre:2006,Rabl:2006,WQS:2008,Petrosyan:2008,Tordrup:2008}.
The small effective mode volume together with the long photon
lifetime allow a strong coupling.

The superconducting qubit to CPWR coupling has been implemented
and studied by several groups \cite{Blais:2004,
Wallraff:2004,Sillanpaa:2007,Hofheinz:2008,Fink:2008}. In this
letter we concentrate on the magnetic coupling of a microwave
photon in a CPWR to a collective hyperfine qubit in an ensemble of
ultracold atoms. We show below that even though the  magnetic
coupling strength is much weaker than the optical dipole coupling,
one can achieve strong coupling with currently available
technology of circuit cavity quantum electro dynamics and
ultracold atomic ensembles on an atomchip.

As particular qubit example we consider a hyperfine transition in
$^{87}$Rb between $|F=2,m_{F}\rangle$ and $|F=1,m_{F}\rangle$
states which frequency of $6.83$ GHz being ideally suited for a
CPWR. In principle both systems can be integrated in a hybrid
device on an single superconducting atomchip
\cite{Nirrengarten:2006,Mukai:2007}. Besides the transfer of a
single photon to the atomic ensemble as a quantum memory and back,
such a hybrid quantum system opens up many different other
possibilities.  For example nondestructive microwave detection of
the atomic density will allow to continuously monitor BEC
formation or changing operating parameters one can achieve a
superradiant microwave source, a micro maser. Optically pumped
atoms are a heat bath close to $T=0$ and will strongly suppress
thermal photons in the coupled resonator mode.  Adiabatic
microwave potentials will allow to couple the quantum properties
of the resonator mode to the mechanical motion of the atoms.

\begin{figure}[t]
  \centerline{\includegraphics[width=\columnwidth,keepaspectratio]{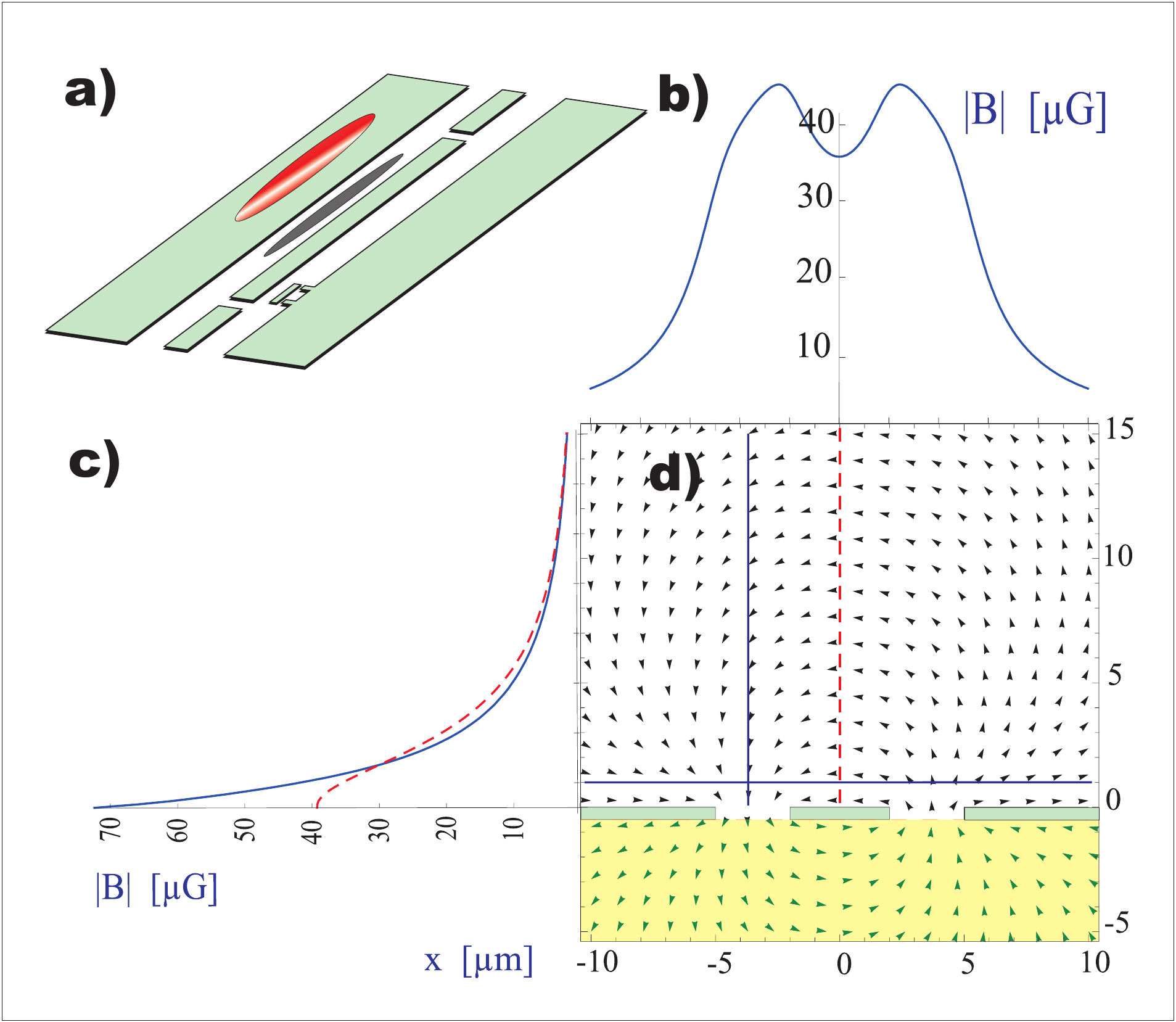}}
  \caption[CPW]{(color online) A CPWR including a solid-state qubit and a cloud of
  ultracold atoms trapped above one of the gaps. Vector plot of the
  magnetic field in the resonator (d). The resulting magnetic field
  strength of a single photon as a function of lateral distance 1 $\mu$m
  above the chip (b) and as a function of distance to the chip over
  the gap (full line) and over the central conductor (dashed line)
  (c). } \label{fig:CPW}
\end{figure}

The coplanar wave guide resonators developed for circuit cavity
QED consist of three conducting stripes: the central conductor
plus two ground planes (Fig.~\ref{fig:CPW}). Their electromagnetic
field is strongly confined near the gap between conductor and the
ground planes. Using atomchip technology
\cite{Folman:2002,Fortagh:2007} large ensembles of ultracold atoms
exceeding $N>10^5$ can be positioned only a few $\mu$m above this
gap \cite{Lin:2004,Aigner:2008}, where they experience the very
strong localized magnetic field of the CPW-mode. The high
concentration of field energy near the surface results in a
dramatic reduction of the effective volume $V_\mathrm{eff} \sim
\frac{\pi}{2} \; \lambda \; l^2$  of the resonator mode. For the
$^{87}$Rb microwave transition at 6.83~GHz (wavelength $\lambda
\sim 3$ cm) and a typical decay length $l \sim3 \mu m$ of the
field of the order of the gap size $W$, one expects an enhancement
of the atom-photon coupling strength of ${(\lambda / l)}\sim
10000$ for ($W=3\,\mu$m). A full calculation of the local
electromagnetic field \cite{Simons:1982,Collin:2001} as shown in
in Fig.~\ref{fig:CPW} confirms these estimates \cite{FullCalc},
and we obtain at the location of the atoms a fields of $>40 \;
\mu$G for a single photon.

For an detailed treatment of the coupling we write the single mode
electromagnetic field operators as:
\begin{eqnarray}\label{eq:fieldoperator}
\vec{E}^{\gamma}(\vec{r},t)&=&\frac{\vec{e}_{\text{tr}}^{\hspace{0.5mm}\gamma}(x,y)}{\sqrt{2}}
\left(a_{\gamma}e^{i(\gamma z-\omega _{\gamma}t)}+a_{\gamma}^{\dagger}e^{-i(\gamma z-\omega_{\gamma}t)}\right)\nonumber\\
\vec{B}^{\gamma}(\vec{r},t)&=&i\frac{\vec{b}_{\text{tr}}^{\hspace{0.5mm}\gamma}(x,y)}{\sqrt{2}}
\left(a_{\gamma}e^{i(\gamma z-\omega _{\gamma}t)}-a_{\gamma}^{\dagger}e^{-i(\gamma z-\omega_{\gamma}t)}\right)\, ,\nonumber
\end{eqnarray}
where $a_{\gamma}^{\dagger}$ and $a_{\gamma}$ represent the boson
creation and destruction operators for the microwave photons.
$\omega=2\pi\nu$ is the angular frequency of the microwave and
$k_{\gamma}={2\pi}/{\lambda}$ is the propagation constant with
wavelength $\lambda={c}/{\sqrt{\epsilon_\mathrm{eff}}}$. The
effective relative dielectric constant $\epsilon_\mathrm{eff}$ has a
value between the substrate value and 1 (vacuum) and depends on
the actual dimensions of the CPW \cite{Wadell:1991}.

The corresponding mode functions
$\vec{e}_{\text{tr}}^{\hspace{0.5mm}\gamma}(x,y)$ and $
\vec{b}_{\text{tr}}^{\hspace{0.5mm}\gamma}(x,y)$ are strongly
varying in space depending on the CPW geometry and have to be
determined numerically. To satisfy the proper field commutators
they have to be normalized to:
\begin{equation}\label{eq:normalization}
\frac{1}{2}\int dV
\epsilon(\vec{r}\hspace{0.5mm})\left|\vec{e}_{\text{tr}}\right|^2=\frac{1}{2\mu_0}\int
dV \left|\vec{b}_{\text{tr}}\right|^2=\frac{1}{2}\hbar\omega
_{\gamma} .
\end{equation}
They represent the field amplitude per photon.  Note that the
permittivity has to be included in the integral.  In the following
we assumed the substrate to be non magnetic. The field Hamiltonian
then reads:
$\mathcal{H}_\gamma=\hbar\omega_{\gamma}\left(a_{\gamma}^{\dagger}a_{\gamma}+\frac{1}{2}\right)$,
where $\omega_{\gamma}$ is the cavity resonance frequency.

For a ground state Rb atom the dominant interaction with a
microwave field are the M1-dipole transitions between the atomic
hyperfine states
$|F=2,m_{F}\rangle\leftrightarrow|F=1,m_{F}^{\prime}\rangle.$ This
leads to the interaction Hamiltonian:
\begin{equation} \label{mu_int}
\mathcal{H}_{int}= \vec{\mu} \cdot\vec{B}^{\gamma}
=\frac{\mu _B}{\hbar}\left(g_S\vec{S}-\frac{\mu_N}{\mu_B}g_I\vec{I}\right)\cdot\vec{B}^{\gamma}
\end{equation}

Assuming an external bias field $\vec{B}_0$ as quantization axis
for the  atomic magnetic moment, transitions driven by a
transverse field $\vec{B}^{\gamma}\perp\vec{B}_0$ follow the
selection rules $\Delta m_{F}=m_{F}-m_{F}^{\prime}=\pm 1$.
Longitudinal fields $\vec{B}^{\gamma}\parallel\vec{B}_0$ induce
$\Delta m_{F}=0$ transitions.

The transverse fields generated by the quasi TEM mode of the CPWR
cavity couples therefore to the two transitions $\Delta m_{F}=1$
and $\Delta m_{F}=-1$. By adjusting the Zeeman splitting of the
hyperfine states via a typical longitudinal Ioffe bias field of
$B_0\sim 1$ Gauss we can ensure that the CPW-mode is only resonant
with one of those two transitions. Thus the atom can be modelled
effectively by a two level system and we simply denote the two
coupled atomic states by $|2\rangle=|F=2,m_{F}\rangle$ and
$|1\rangle=|F=1,m_{F}^{\prime}\rangle$ with energies $E_2$ and
$E_1$.

For the internal dynamics of an ensemble of $N$ atoms, we thus get
an effective Hamiltonian in a standard Jaynes-Cummings form
\cite{Knight:1993}:
$$H_{atom}=\sum_i\hbar\omega_a \, \hat{\pi}_i^{\dagger}\hat{\pi}_i-
\sum_i\left\{\hat{\mu}_i\cdot\hat{B}(\vec{r}_i)+
\hat{\mu}_i^{\ast}\cdot\hat{B}^{\dagger}(\vec{r}_i)\right\}.$$
Here $\omega_a$ is the atomic transition frequency and
$\hat{\pi}_i^{\dagger}$ is the excitation operator of the $i^{th}$
atom and $\hat{\mu}_i=\vec{\mu}_i^{\ast}\ \hat{\pi}_i+\vec{\mu}_i
\hat{\pi}_i^{\dagger}$ with $\vec{\mu}_i$ is the transition matrix
element of the magnetic dipole moment as defined in Eq.
(\ref{mu_int}).

Typically an ensemble of ultracold Rb atoms is confined in an
elongated potential on the atomchip and has a transverse extension
of $d < 1 \mu$m and a length of up to a few mm, much smaller then
the transverse variation of the magnitude, respectively the
wavelength of the microwave field. Positioning the atomic ensemble
longitudinally in vicinity of the field maximum we can, to a first
approximation, neglect in the change of the magnetic field along
the cloud in magnitude and direction and fix
$\vec{b}^{\gamma}_{trans}(r_i) \approx
\vec{b}^{\gamma}_{trans}(\bar{R})$ for all atoms. Here $\bar{R}$
is the mean transverse position. This allows to define a simple
collective atomic excitation operator of the form
$\tilde{\pi}=\frac{1}{\sqrt{N}}\sum_i\hat{\pi}_i$ to rewrite the
interaction Hamiltonian in the Tavis Cummings form
\cite{Tavis:1968}:
\begin{equation}
\label{eq:TCHam}
H=\hbar\omega_{\gamma} a^{\dagger} _{\gamma} a_{\gamma}+
  \hbar\omega_a \tilde{\pi}^{\dagger}\tilde{\pi}+
  \hbar {g}_\mathrm{eff}\tilde{\pi}^{\dagger}a_{\gamma}+
  \hbar{g}_\mathrm{eff}^{\ast} a^{\dagger}_{\gamma}\tilde{\pi},
\end{equation}
where ${g}_\mathrm{eff}=\sqrt{N}\ g$ and $\hbar
g=\frac{1}{\sqrt{2}}(\vec{b}^{\gamma}_{trans}(\bar{R})\cdot\vec{\mu})$.
From this we can immediately read off the effective coupling
strength $g_\mathrm{eff}$ for the first symmetric atomic excitation,
which is enhanced by a factor of $\sqrt{N}$.

As a concrete example we obtain a matrix element of $ 0.86 \mu_B$
for a $m_F=2$ to $m_F=1$ transition in the $^{87}$Rb ground state.
Taking into account the calculated field of a CPWR (see fig.
\ref{fig:CPW}) we obtain a single photon - single atom Rabi
frequency of typically $g/2\pi \sim 40$ Hz at a height of a few
$\mu$m. For an atomic ensemble of $N \sim 10^6$ $^{87}$Rb atoms
coherent collective coupling $
g_\mathrm{eff}/2\pi=\sqrt{N}g/2\pi\sim 40$~kHz, dominates over
cavity decay $\kappa/2\pi = \nu/Q \sim 7$~kHz and one would get
several exchanges between a microwave photon in he cavity and a
collective atomic excitation before the photon decays.

The Hamiltonian (Eq.~\ref{eq:TCHam}) can be diagonalized by
eigenstates forming a weighted coherent superposition of
collective atomic excitations and a photon depending on system
parameters. Controlling the relative weights of the superposition
via atomic or cavity tuning adiabatically switches excitation
between the microwave and the atomic qubit. The strong coupling
regime allows to perform this transfer fast enough to avoid
decoherence.

The upper collective atomic qubit state is a delocalized symmetric
superposition of all possible single atom excitations and one can
in principle achieve long coherence times for the collective
atomic hyperfine qubit. Decay rates of $\gamma/2\pi \sim 0.3$~Hz
were recently demonstrated for hyperfine excitations
\cite{Treutlein:2004}. As demonstrated in atom ensemble
experiments \cite{Lukin:2003} such an excitation can be
efficiently read out by forward coherent Raman scattering into a
travelling wave optical photon.  This will allow to complete the
transfer from solid-state qubits via an atomic quantum memory to
photons as flying qubits.

Besides implementing a quantum interconnect between solid-state
qubits, atoms and even photons, our system offers many further
interesting possibilities, which we will discuss in short examples
below.

\begin{figure}[t]
  \centerline{\includegraphics[width=\columnwidth,keepaspectratio]{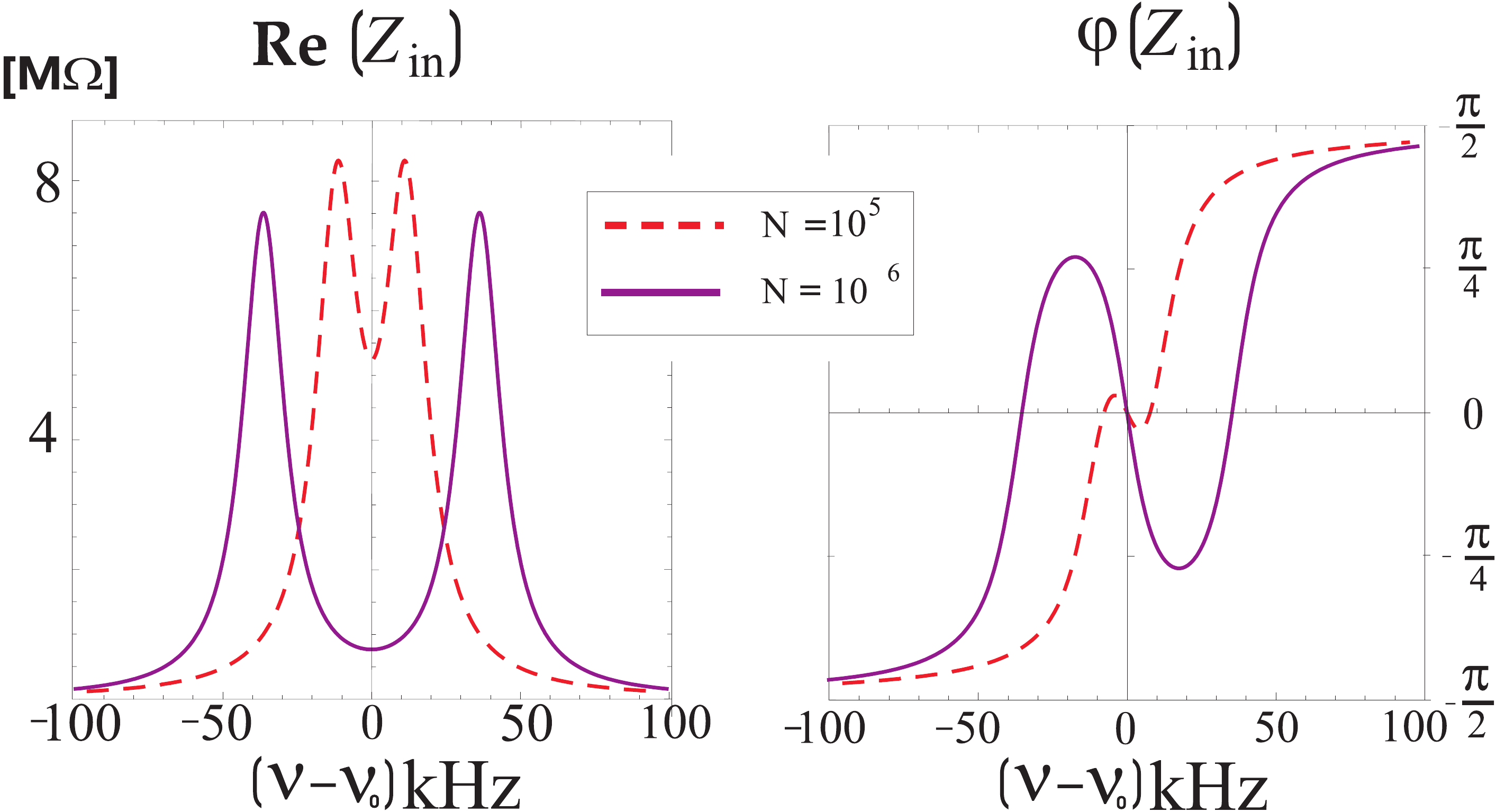}}
  \caption{(color online) Response of the atom - cavity system as illustrated by its complex impedance Z:
  \emph{(left)} The real part of Z is related to the spectrum and \emph{(right)}
  the phase of Z directly illustrates the phase shift of the transmitted microwave radiation.
  The calculations are for Q=$10^6$, N=$10^6$ and N=$10^5$
  and plotted vs. incident field detuning from the resonance frequency.
  }
\label{Fig:Reflection}
\end{figure}

Whenever the effective Rabi splitting $g_\mathrm{eff}$ is larger
then the cavity line width $\kappa$, it manifests itself in the
spectrum of the transmitted and reflected fields, exhibiting
resonances at these collective excitations (see
Fig.~\ref{Fig:Reflection}). Hence one can use the microwave field
as nondestructive probe for the integrated atomic density in the
mode.

Let us note that, as compared to standard cavity QED here the line
width of the atomic excitation is not limited by spontaneous decay
as for all practical purposes both hyperfine states can be
regarded as stable. Hence the decay rate it will effectively be
given by non radiative losses such as the lifetime of the atoms in
the trap, which can be in the order of seconds. In view of the
large difference of atomic and cavity decay the single atom
cooperativity parameter $C=g^2/( \kappa \gamma)$, which can reach
$C \sim 1$, is only partly meaningful. While a single excited atom
will still emit one photon predominantly into the cavity mode, the
large cavity line width prevents direct single atom detection via
resonator transmission.

In the microwave regime the transmitted field amplitude and phase
are directly accessible. The corresponding phase shift of the
transmitted field is plotted in Fig.~\ref{Fig:Reflection}b, for an
empty cavity and one filled with $10^6$ atoms. Note that measuring
the phase shift will not only be a sensitive probe of the atom
number but, as the phase will change sign when the atoms are
transferred to the upper hyperfine state, it can also be used for
preparation and readout of spin states.

One important aspect neglected so far are thermal photons in the
resonator. The number of photons in the cavity is given by
$\bar{n}_T = \exp(-{\hbar \omega}/{k_B T})$. With 6.83~GHz
corresponding to a temperature $T \sim 350$~mK cooling to below
100 mK is required to have an empty cavity ($\bar{n}_T < 0.1$).
However a perfectly polarized BEC with all atoms in the lower
hyperfine state has a very low effective internal temperature. A
relative purity in the polarization of 10$^{-5}$ corresponds to
${\hbar \omega_\gamma}/{k_B T}=\exp(10^{-5})=11.5$ or a
temperature of $T \sim 30$~mK. Hence the coupling of the two
systems can lead to an energy flow towards the ensemble of
ultracold atoms. We estimate the photon absorption rate from the
cavity into the atomic ensemble to ${\gamma_{c}}/{2 \pi} \sim {g^2
N}/(\gamma_a\:2\pi) \sim 8.6$~MHz, assuming an upper state with a
lifetime of $\gamma_a^{-1} \sim$~1~ms (for this experiment a state
that is very short lived is used) which has to be compared to the
heating rate $R \sim \kappa\bar{n}_T$. The suppression of the
thermal photons is then given by ${\kappa}/({\gamma_c + \kappa})$.
We can remove thermal photons from the mode as long as
$\gamma_{c}\gg\kappa$, which can be for several 10 seconds.

Superradiance from a completely inverted atomic ensemble has been
first discussed in the microwave context \cite{Dicke:1954} and
lead to extensive theoretical and experimental studies
\cite{Gross:1982}. Here we could study its magnetic analogue in a
very clean form by preparing an almost perfectly inverted atomic
system with all atoms in $F=2$. This situation is very close to
the original model for a superradiant system proposed by Dicke
\cite{Dicke:1954}, where a highly excited atomic system is
supposed to spontaneously emits coherent multi-photon pulses.
Here, with the spontaneous life time of the excited state
practically infinite, we can assume the dominant decay to happen
solely via the cavity mode and we get an emission rate of $ N
g^2/\kappa \sim 5\cdot 10^5 \mathrm{s}^{-1}$ at which about $10^6$
photons are collectively emitted into the microwave mode in a
coherent pulse.

The field in the cavity mode can exert forces onto the cloud of
ultracold atoms through microwave induced dressed state potentials
(microwave ac-Stark shift) \cite{Muskat:1987,Agosta:1989}. A
coupling strength of $g /{2 \pi} \sim 40$~Hz results in small
modifications (dressed state shifts) of the trapping potential and
forces are small on the few photon level. Potential energies in
the order of a typical chemical potential of a trapped 1d cloud $V
\sim 1 kHz$ appear for microwave fields of $1000$ photons in the
mode, which still correspond to a minute powers of only $P \sim 5
\times 10^{-16} \hbar \omega \sim 10^{-16}$ W.  In the quantum
noise of the cavity field, and its time evolution will have a
strong influence on the atomic motion in the trap. In addition a
microwave power in the nano Watt regime will create very large
forces and could help for atomic positioning towards optimum
coupling.

Comparing with other related CQED systems, the cooperativity for
the collective qubit state is very large and even comes close to
the values for a BEC coupled to an resonator on a strong optical
transition \cite{Slama:2007,Brennecke:2007,Colombe:2007}. Note
that the involved transition frequency is much lower in our case.
Hence one could even envisage reaching the regime of the quantum
phase transition to a collective superradiant phase, predicted for
$g N \approx \omega_c$ in a classic paper by Hepp
\cite{Hepp:1973}.

In conclusion we found that coherent strong coupling between a
collective spin Dicke state, as it is used for ensemble qubits,
and a microwave photon from a CPW resonator is feasible by
combining current state of the art technology of atomchips and
superconducting microwave resonators. Such a \emph{quantum
interconnect} will allow to transfer a quantum state of a
solid-state qubit into the atomic ensemble and store it there as
interface for long distance quantum communication. In addition we
get a microwave realization of the Tavis Cummings model, where
super radiance can be studied and maybe even used for cooling and
and generating strong coherent light forces with microwaves.

This work was supported by the European Union project MIDAS and
the Austrian FWF projects P17709 and Z118-N16, J.V. acknowledges
support from the Marie Curie Fellowship MIEDFAM. We thank M.
Trinker and A. Wallraff for stimulating discussions.


\end{document}